# Asymmetrical Filtering Impairments Mitigation for Digital-Subcarrier-Multiplexing Transmissions Enabled by Multiplication-free K-State Reserved Complex MLSE

Hexun Jiang[(1)], Zhuo Wang[(1)], Chengbo Li[(1)], Weiqin Zhou[(1)], Shuai Wei[(1)], Yicong Tu[(1)], Heng Zhang[(1)], Wenjing Yu[(1)], Yongben Wang[(1)], Yong Chen[(1)], Ye Zhao[(2)], Da Hu[(2)], Lei Shi[(2)]

[(1)] ZTE cooperation, Shenzhen, China, jiang.hexun@zte.com.cn
[(2)] Sanechips Technology CO.,LTD, Shenzhen, China

**Abstract** *We propose a multiplication-free K-state reserved complex maximum-likelihood-sequence-estimation (MLSE) to mitigate asymmetrical filtering impairments in digital-subcarrier-multiplexing transmissions. A required optical-to-noise ratio of 1.63 dB over the conventional real MLSE is obtained after transmitting 90 GBaud DSCM DP-16QAM signal over 14 WSSs without multiplications.* ©2025 The Author(s)

## Introduction

The optical network is evolving towards higher rates and enhanced flexibility. As the single-wavelength data rate increases from 400 G to 800 G and 1.6 T, fiber nonlinearity and equalization enhanced phase noise (EEPN) become the primary link induced impairments. In this context, digital subcarrier multiplexing (DSCM) systems are attracting widespread attentions with greater tolerance to fiber nonlinearity and EEPN, in comparison with the single-carrier systems by the use of low baud-rate subcarriers [1]. With the advancement of elastic optical networks, in-line filtering impairments become more and more severe, due to the widespread deployment of reconfigurable optical add-drop multiplexers (ROADM) based on wavelength-selective switches (WSSs) and the reduction of channel spacing resulting from limited spectral resources [2]. Unfortunately, DSCM systems are prone to optical filtering impairments, because edge subcarriers suffer significant and asymmetrical filtering, while the filtering impairments affecting center subcarriers are typically negligible. In practical systems, the central frequency of transmitter (Tx) commercial external cavity lasers (ECLs) typically drifts by up to $\pm 1.5$ GHz over their operational lifetime, which exacerbates filtering impairments due to a center-frequency mismatch between the Tx lasers and the WSSs.

Conventional feed-forward equalizer (FFE) are commonly employed to compensate for the filtering impairments. However, FFE can also amplify noise in regions of high-frequency attenuation, resulting in poor performance under severe filtering conditions. In the intensity modulation and direct detection system, a maximum likelihood sequence estimation (MLSE) combined with a post-filter can effectively mitigate the equalizer-enhanced noise [3]. This approach can also be applied in coherent single-carrier systems with two real MLSEs to handle real and imaginary parts of complex signal, respectively. However, for edge subcarriers in DSCM, they undergo asymmetrical filtering, leading to crosstalk between the real and imaginary components. As a result, using two real MLSEs fails to compensate for the crosstalk, leading to sub-optimal performance. On the other hand, complex MLSE has a much larger signal constellation set, resulting in a substantially increased number of branch transitions in the trellis diagram. Thus, it becomes too computationally intensive for practical implementation, even though employing some low-complexity implementation [4,5].

In this paper, we propose a K-state reserved complex MLSE to mitigate asymmetrical filtering impairments. We first present the derivation of the complex post-filter calculation. Then, the principle of the proposed reserved state scheme is given, which can reduce the computational complexity significantly. The effectiveness of the proposed algorithm is experimentally verified by transmitting 4-subcarrier 16-quadrature amplitude modulation (16-QAM) DSCM signals with an aggregated baud rate of 90 GBaud through 14 WSSs with 100 G channel spacing. The proposed MLSE achieves 1.63 dB gain in required optical signal-to-noise ratio (ROSNR) compared to the conventional real MLSE in a multiplication-free manner.

## Principle of K-state reserved complex MLSE

The signal $y(n)$ after FFE can be expressed as
$$y(n) = x(n) + N(n), \quad (1)$$
where $x(n)$ and $N(n)$ represent transmitted symbols and noise after FFE, respectively. The signal $z(n)$ after applying the post-filter $h$ can be expressed as
$$z(n) = \sum_{i=0}^{L-1} y(n-i)h(i)$$
$$= \sum_{i=0}^{L-1} x(n-i)h(i) + \widehat{N}(n), \quad (2)$$

where $\widehat{N}(n)$ is the noise after the post-filter. The iterative formula of $h$ can be derived using the gradient descent method like [6]

$$h_k(i) = h_{k-1}(i) - \mu\widehat{N}(k)N^*(k-i), \quad (3)$$

where * represents the conjugate operation. Subsequently, a standard complex MLSE implemented by the Viterbi algorithm is applied to eliminate the inter-symbol interference (ISI) introduced by the post-filter. In practice, the coefficients $h$ could be precomputed using a known training sequence, allowing this part of the computational complexity to be omitted.

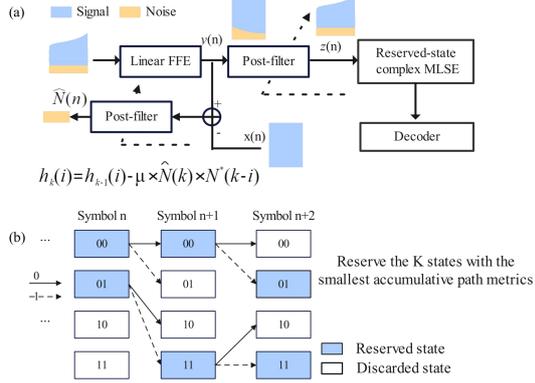

**Fig.1** (a) Complex MLSE process. (b) The trellis diagram in the proposed K-state reserved MLSE.

To reduce the implementation complexity of complex MLSE, we reserve only the K states with the smallest accumulative path metrics, which can reduce the branch transitions in the trellis diagram from $M^L$ to KM. The principle is illustrated in Fig. 1 (b), where $M = 2, L = 3$ and $K = 2$. 2 states (00&11) are reserved for the calculation of the branch transitions for symbol n+1, and only 4 rather than 8 branch transitions need to be calculated. On the other hand, branch transitions are generally calculated by the Euclidean distance (ED):

$$\left|z(n) - \sum_{i=0}^{L-1} tx(n-i)h(i)\right| \quad (4)$$

In practice, $\sum_{i=0}^{L-1} tx(n-i)h(i)$ can be pre-calculated and stored in a lookup-table (LUT) $l(n)$, and Eq. (4) can be changed into

$$|z(n) - l(n)|. \quad (5)$$

For the real symbols, ED can be calculated without multiplication but it requires two real multiplications for the complex symbol. To reduce the complexity further, the ED in Eq. (5) can be substituted by the right-angled sides distance (RASD) as follows:

$$|Re(z(n)) - Re(l(n))| + |Im(z(n)) - Im(l(n))|, \quad (6)$$

where only two real additions are required. Table 1 shows the branch transitions, multiplications, additions and storage units for the various MLSEs using LUT, where $M$ is the modulation order for complex signals.

**Experimental setup and DSP flow**

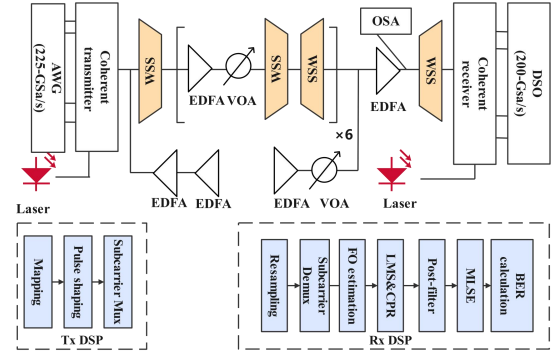

**Fig. 2:** Experimental setup and DSP flow.

Fig. 2 depicts the experimental setup and digital signal processing (DSP) flow. In the Tx DSP, four 22.5 GBaud 16-QAM subcarriers are generated by a root raised cosine filter with a roll-factor of 0.01. Then, the 90 GBaud multiplexed electrical signals are uploaded by a 225 GSa/s arbitrary waveform generation (AWG). Then, a 58 GHz dual-polarization in-phase and quadrature modulator is employed to modulate the optical carrier from the ECL, whose center frequency ranges from 193.6985 THz to 193.7000 THz to emulate a 1.5 GHz laser frequency drift observed in practical systems. The signal channel, along with dummy channels generated by two Erbium-doped fiber amplifiers (EDFAs), is multiplexed via a WSS to emulate a wavelength-division multiplexing system. Then, 12 WSSs, along with EDFAs and variable optical attenuators (VOAs), are used for filtering the optical signal while maintaining constant output power. The channel spacing and center frequency of all WSSs is set to 100 GHz and 193.7000 THz, respectively. Additional noise is injected at the end of the link to control the OSNR. The in-band OSNR after optical filtering is accurately measured using the channel turn-off method. At the receiver (Rx) side, a 70 GHz coherent receiver, using a similar laser with center frequency fixed at 193.7000 THz, is utilized to detect the optical signal. The electrical waveform is captured by a 59 GHz real-time digital oscilloscope (DSO, Tek DPO70000SX) at 200 GSa/s sampling rate. For the Rx offline processing, 4 subcarriers are resampled, demultiplexed and then processed separately. After FO estimation, a least-mean-square based FFE embedded with carrier phase recovery (CPR) is applied. Then, the proposed post-filter and MLSE schemes are executed. Finally, the mean bit error rate (BER) across the 4 subcarriers is obtained.

**Tab. 1:** Complexity comparison.

|  | Two Real MLSEs | Standard Complex MLSE | Simplified ED MLSE | Simplified RASD MLSE |
|---|---|---|---|---|
| Branch transitions | $2 \times M^{L/2}$(128) | $M^L$(4096) | KM(128) | KM(128) |
| Multiplications | 0 | $2 \times M^L$(8192) | $2 \times$KM(256) | 0 |
| Additions | $2 \times M^{L/2}$(128) | $M^L$(4096) | KM(128) | $3 \times$KM(384) |
| Storage units | $2 \times M^{L/2}$(128) | $M^L$(4096) | $M^L$(4096) | $M^L$(4096) |

**Experimental result and discussion**

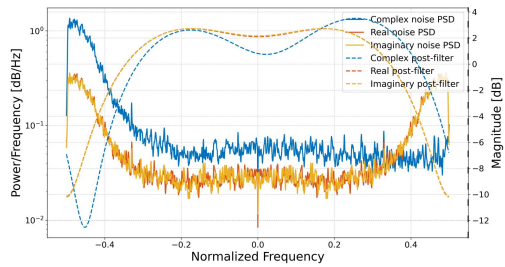

**Fig.3:** Complex and real noise spectrum and frequency response of the complex and real post-filter of the subcarrier with the worst performance when Tx FO is -1.5GHz

Fig.3 shows the noise spectrum and frequency response of post-filter of the subcarrier with the worst performance when the Tx FO is -1.5 GHz. The complex noise spectrum shows a higher amplitude on the left side compared to the right whereas the real and imaginary noise spectra remain inherently symmetrical. The complex post-filter presents an asymmetrical frequency response, effectively suppressing the noise spectrum on the left side.

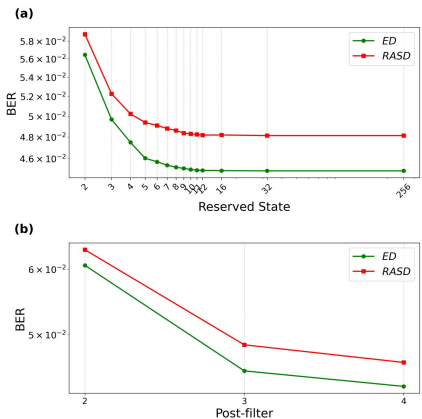

**Fig.4:** (a) BER versus (a) the reserved state and (b) post-filter length of the worst subcarrier when the Tx FO is -1.5 GHz.

Then, the impacts of the number of reserved states and the post-filter length are investigated. First, the channel length is fixed at 3. As the number of reserved states increases, the performance improves and nearly saturates at 8, as shown in Fig. 4(a). Subsequently, the number of reserved states is fixed at 8, and the channel length is varied from 2 to 4, with the corresponding results depicted in Fig. 4(b). A longer post-filter yields better performance due to the finer frequency resolution and more effective noise suppression. Considering the trade-off between complexity and performance, a post-filter length of 3 and 8 reserved states are adopted in subsequent experiments.

For the purpose of performance comparison, we focus on the following DSP cases: 1) FFE; 2) standard real MLSE; 3) standard complex MLSE; 4) simplified ED MLSE; 5) simplified RASD MLSE. As shown in Fig.5 (a), only FFE could not reach BER threshold of 2.2e-2. All the complex MLSEs achieve nearly identical performance and outperform the real MLSE. Fig. 5(b) shows that the performance degrades as the Tx FO increases, primarily due to more severe filtering impairments. When the Tx FO reaches -1.5 GHz, the simplified ED MLSE and simplified RASD MLSE achieve ROSNR gains of 1.76 dB and 1.63 dB over the real MLSE, respectively. The performance of simplified ED MLSE is nearly the same with standard complex MLSE. Finally, the complexity and storage requirements in our experimental DSP cases are summarized in brackets as shown in Table I. At the cost of additional storage units and some additions, the simplified RASD MLSE does not require multiplications like real MLSE.

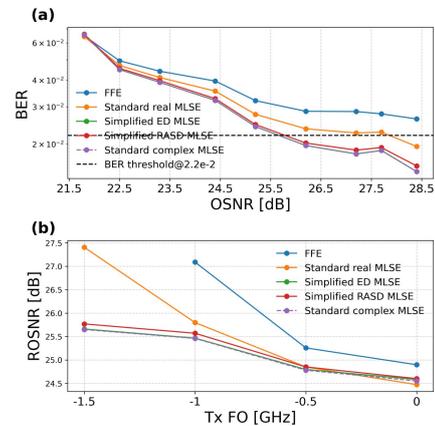

**Fig. 5:** (a) BER versus OSNR at -1.5GHz Tx FO. (b) ROSNR versus Tx FO.

**Conclusions**

In summary, a K-state reserved complex MLSE is proposed to mitigate asymmetrical filtering impairments. We only reserve *K* states in the trellis diagram, reducing the computational complexity sharply. In the experiment, the simplified RASD MLSE achieves more than 1.6 dB ROSNR improvement over real MLSEs without multiplications. The proposed scheme offers a cost-effective solution for mitigating asymmetrical filtering impairments in future long-haul systems.